%% file: bognar.tex
\documentclass[mypaper,7pt,twoside]{CoAst}
\usepackage{epsf,graphicx,fancyhdr}
\input{CoAst_layo}

\begin{document}
\sf

\chapterDSSN{GD 99 - an unusual, rarely observed DAV white dwarf}{Zs. Bogn\'ar, M. Papar\'o, B. Steininger and  G. Vir\'aghalmy}

\Authors{Zs. Bogn\'ar$^1$, M. Papar\'o$^1$, B. Steininger$^2$, G. Vir\'aghalmy$^1$} 
\Address{$^1$ Konkoly Observatory, P.O.Box 67., H-1525 Budapest, Hungary\\
$^2$ Institut f\"ur Astronomie, T\"urkenschanzstrasse 17, 1180 Vienna, Austria}

\noindent
\begin{abstract}
New observation of GD 99 was analysed. The unusual pulsation behaviour, showing both long and short periods, has been confirmed. All the available periods show a grouping of short and long period modes with slightly regular spacing. If we interpret the groups separately, a binary nature can be a possible explanation as in similar cases of WD 2350-0054 and G29-38.
\end{abstract}
\\
\\
\indent
GD 99 was previously observed in 1975 (\mbox{McGraw} \& Robinson 1976) and in 2003 (Chynoweth et al. 2004).
 
On three consecutive nights 23-hour-observations were obtained in white light at Piszk\'estet\H o, the mountain station of Konkoly Observatory in February 2002, with a CCD attached to the 1m telescope. Data reduction and frequency analysis were carried out by using the standard IRAF packages and the MUFRAN package (Koll\'ath 1990).

\figureDSSN{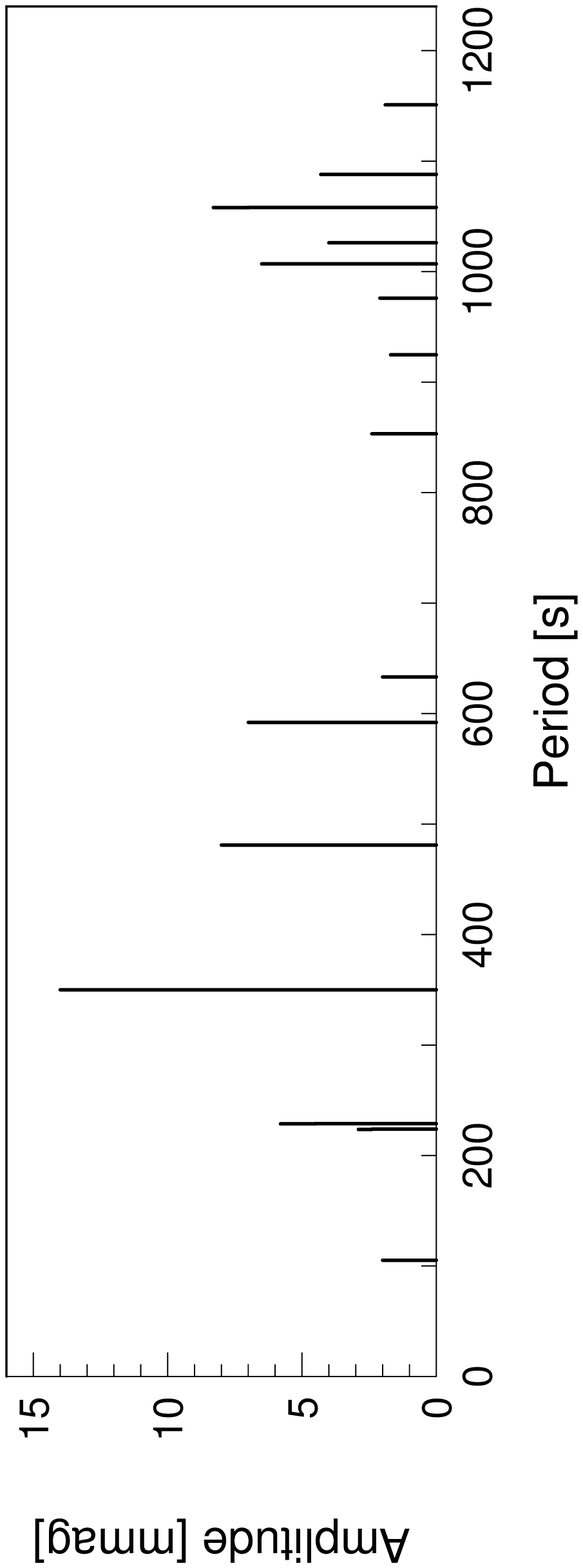}{All the excited periods of GD 99 at different epochs.}{fig1}{!!!ht}{clip,angle=-90,width=105mm}

Four periods (1058.1, 228.7, 1026.1 and 223.9 s) were identified with amplitudes (7.0, 5.8, 4.0 and 2.4 mmag) in 2002. Three of them (1058.0/8.3, 228.9/4.5 and 223.6/2.9, P(s)/A(mma)) are confirmations of modes given by Mukadam et al. (2006) (hereafter M06). The fourth period is a newly identified mode. Their other modes and three short period modes (frequencies given by Bradley (2000); amplitudes given by Clemens (1993)) are presented in Fig~\ref{fig1}. The last three modes are not included in M06 because of uncertainty. The unusual behaviour of \mbox{GD 99} ($T_{eff} = 11820$ K, log $g \sim8.08$) is obvious: despite the well-established trend of decreasing pulsation period with increasing effective temperature, \mbox{GD 99} (situated on the blue edge of the DAV instability strip) shows both short and long periods. Some kind of grouping can be seen both among the long and short period modes. The spacing in the short period group is about twice as large as than the spacing in the long period group.

Recently a new classification criterion was published (M06). Based on the weighted mean period hot, intermediate and cool subclasses were introduced. According to the spectroscopic temperature \mbox{GD 99} belongs to the hot subclass. The weighted mean period puts it into the intermediate class. \mbox{GD 99} could be situated in the cool subclass if we regard only the modes given by M06. It is quite improbable that a single star belongs to each subclasses. If we interpret the group of long and short period modes separately, a plausible explanation could be a binary nature. One component is situated at the hot while the other at the cool boarder of the instability strip. The effective temperature and pulsation period of WD 2350-0054 (Mukadam et al. 2004) and G29-38 (Kleinman 1995) also do not fit the general trend of DAV stars. According to the binary concept, one component of WD 2350-0054 would pulsate, and the other should pass the red edge of the DAV instability strip. In the hypothetic binary concept of G29-38, one component pulsates with a long period, while the other has not passed over the blue edge of the DAV instability strip.

GD 99 definitely needs a more complex investigation (DARC/WET run).


\References{
Bradley, P. A. 2000, BaltA 9, 485\\
Clemens, J. C. 1993, BaltA 2, 407\\
Clynoweth, K. M. et al. 2004, BAAS 36, 1514\\
Kleinman, S. J. 1995, Ph.D. Thesis, Univ. of Texas at Austin\\
Koll\'ath, Z. 1990, Occ. Techn. Notes Konkoly Obs., No. 1\\
McGraw, J. T., Robinson, E. L. 1976, ApJ 205, L155\\
Mukadam, A. S. et al. 2004, ApJ, 612, 1052\\
Mukadam, A. S. et al. 2006, ApJ, 640, 956
}

\end{document}

%% file: CoAst_layo.tex
\renewcommand{\chaptermark}[1]{\markboth{#1}{}}

\newcommand{\kopf}{\small\itshape Comm. in Asteroseismology\\ Vol. 144, 2003}
\newcommand{\Authors}[1]{\begin{center}\normalsize\bf\sf #1 \end{center}}

\renewcommand{\author}[1]{\begin{center}\normalsize\bf\sf #1 \end{center}}
\newcommand{\Address}[1]{\begin{center}\small\sf #1 \end{center}}

\renewenvironment{abstract}{\section*{Abstract}\normalsize\sf}{}
\newcommand{\References}[1]{\begin{flushleft}{\large References\\}\vspace*{2mm}\small #1 \end{flushleft}}

\newcommand{\chapterDSSN}[2]{\chapter[\sf\normalsize #1\\ \footnotesize \hspace*{5mm}by #2 \sf\normalsize][]{#1\\}\rhead[\fancyplain{}{\sf\footnotesize \center{#1}}]{\fancyplain{}{\sffamily\thepage}}\lhead[\fancyplain{\kopf}{\sffamily\thepage}]{\fancyplain{\kopf}{\sf\footnotesize \center{#2}}}}

\newcommand{\figureDSSN}[5]{\begin{figure}[#4]
\centering
\includegraphics*[#5]{#1}
\caption{#2}
\label{#3}
\end{figure}}

\renewcommand{\chaptername}{}